\documentclass{aa}  

\usepackage{graphicx}
\usepackage{txfonts}
\usepackage{hyperref}
\hypersetup{
    colorlinks=true,
    citecolor=blue,
    linkcolor=blue,
    filecolor=blue,      
    urlcolor=blue,
    pdfpagemode=FullScreen,
    }
\usepackage{multirow}
\usepackage{subfig,graphicx}

\begin{document}

   \title{Photometric redshift-aided classification using ensemble learning}
    \titlerunning{Photometric redshift-aided classification using ensemble learning}

   \author{P.~A.~C. Cunha
          \inst{1,}\inst{2}
          \and
          A. Humphrey\inst{2}
          }

   \institute{Faculdade de Ciências da Universidade do Porto, Rua do Campo de Alegre, 4150-007 Porto, Portugal\\
   \email{pedro.cunha@astro.up.pt}
   \and
        Instituto de Astrofísica e Ciências do Espaço, University of Porto, CAUP, Rua das Estrelas, Porto, 4150-762, Portugal\\
    \email{andrew.humphrey@astro.up.pt}
    }

   \date{Received January 17, 2022; accepted April 21, 2022}

 
   \abstract{ We present \texttt{SHEEP}, a new machine learning approach to the classic problem of astronomical source classification, which 
   combines the outputs from the \texttt{XGBoost}, \texttt{LightGBM}, and \texttt{CatBoost} learning algorithms to create stronger classifiers. A novel step in our pipeline is that prior to performing the classification, \texttt{SHEEP} first estimates photometric redshifts, which are then placed into the 
   data set as an additional feature for classification model training; this
   results in significant improvements in the subsequent classification performance.
    \texttt{SHEEP} contains two distinct classification methodologies: (i) Multi-class and  (ii) one versus all with correction by a meta-learner. 
    We demonstrate the performance of \texttt{SHEEP} for the classification of stars, galaxies, and quasars using a data set composed of SDSS and WISE photometry of 3.5 million astronomical sources. The resulting F1-scores are as follows:   0.992 for galaxies;   0.967 for quasars; and     0.985 for stars. In terms of the F1-scores for the three classes, \texttt{SHEEP} is 
    found to outperform a recent \texttt{RandomForest}-based classification approach using an essentially identical data set. Our methodology also facilitates  model and data set explainability via feature importances; it also allows the selection of sources whose uncertain classifications may make them interesting sources for follow-up observations.}


   \maketitle
%

\section{Introduction}

Imaging and spectroscopic surveys are one of the main resources for the understanding of the baryonic content of the Universe. The data from these surveys enables statistical studies of stars (e.g. \citealt{2014A&A...562A..71B}), quasars (hereafter QSOs; e.g. \citealt{2006ApJS..166..470R}), and galaxies (e.g. \citealt{2003MNRAS.341...33K}), and  the discovery of more peculiar objects such as the previously elusive Type 2 quasars (e.g. \citealt{2013MNRAS.435.3306A, 2016MNRAS.459.3144Z}). 

Wide-area surveys, both ground- and space-based, have yielded high volumes of data, revolutionising the field of astronomy [e.g. Sloan Digital Sky Survey (SDSS): \citealt{1998AJ....116.3040G}; \citealt{2000AJ....120.1579Y}]. Future surveys will continue to give us more detailed imaging, at a range of wavelengths [e.g. the Large Synoptic Survey Telescope (LSST): \citealt{2019ApJ...873..111I}; the Dark Energy Spectroscopic Instrument (DESI): \cite{2019AJ....157..168D}; {\it Euclid}: \citealt{2021arXiv210801201S}; the James Webb Space Telescope: \citealt{2006SSRv..123..485G}].

While high quality spectroscopic observations of sources is desirable, they can be time consuming even with modern multiplexing multi-object spectrographs. Conversely, photometry measured from images allows the efficient assembly of spectral energy distributions (SEDs) for very large samples of objects, thus continues to play an important role in source classification, and in the estimation of redshifts and physical properties (e.g. \citealt{2000ApJ...536..571B}). 

Among the simplest use-cases for photometric SEDs are  single-colour or colour-colour selection techniques, which function as lossy dimensionality reduction methods and allow the separation of some source classes \citep[e.g.][]{1956BOTT....2n...8H,2004ApJ...608..752B}.
While intuitive and usually simple to apply, colour-colour methods are often crude, and typically only use a small subset of the information available in the SED.

On the other hand, spectral template fitting techniques make use of a wider range of features within the SED  to derive various physical properties and to estimate its photometric redshift \cite[e.g.][]{2000A&A...363..476B,Laigle2016}. Although template fitting methods generally outperform colour-colour methods, they can be computationally expensive to apply to very large samples since  the sources are typically  modelled on an individual basis.

Machine learning offers a promising alternative to the colour-colour and template fitting methods, for two main reasons: (i) the full range of photometric information in a source SED can be made use of and  (ii) once a machine learning model is trained, it can be computationally inexpensive to apply it to new samples. While there are various ways in which machine learning can be applied to astronomical data, the three most common applications are  source classification
\citep[e.g.][]{2008AIPC.1082....9E, 2016A&A...592A..25K, 2016A&A...596A..39K, 2019AJ....157....9B, 2020A&A...639A..84C},
physical property estimation 
\citep[e.g.][]{2017MNRAS.465.1144U, 2019A&A...622A.137B, 2019yCat..74861377D, 2019arXiv190508996S, 2021MNRAS.502.2770M}, 
and photometric redshift estimation
\citep[e.g.][]{2018A&A...619A..14F, 2019ascl.soft10014S, 2019NatAs...3..212S, 2021Galax...9...86C, 2021A&A...649A..81N}.

In this work we describe \texttt{SHEEP}, a supervised machine learning pipeline that estimates photometric redshifts and uses this information when subsequently classifying the sources into the galaxy, QSO, and star classes. In Section \ref{data} we describe the data set used in this work. In Section \ref{metrics} we present the main statistical metrics used to evaluate the performance of our pipeline. Section \ref{POC} presents the results from a proof of concept task for object type classification using redshift as a feature. In Section \ref{sheep} we describe the \texttt{SHEEP} pipeline, the feature engineering process, the pipeline architecture and motivations, and   the selection of features. In Section \ref{results} we apply \texttt{SHEEP} to an SDSS and WISE photometric data set, and discuss the performance. Finally, in Section \ref{conclusions} we provide a summary and the  conclusions of this work.

\section{Data}
\label{data}
\begin{figure}[h!]
\centering
  \includegraphics[width=\linewidth]{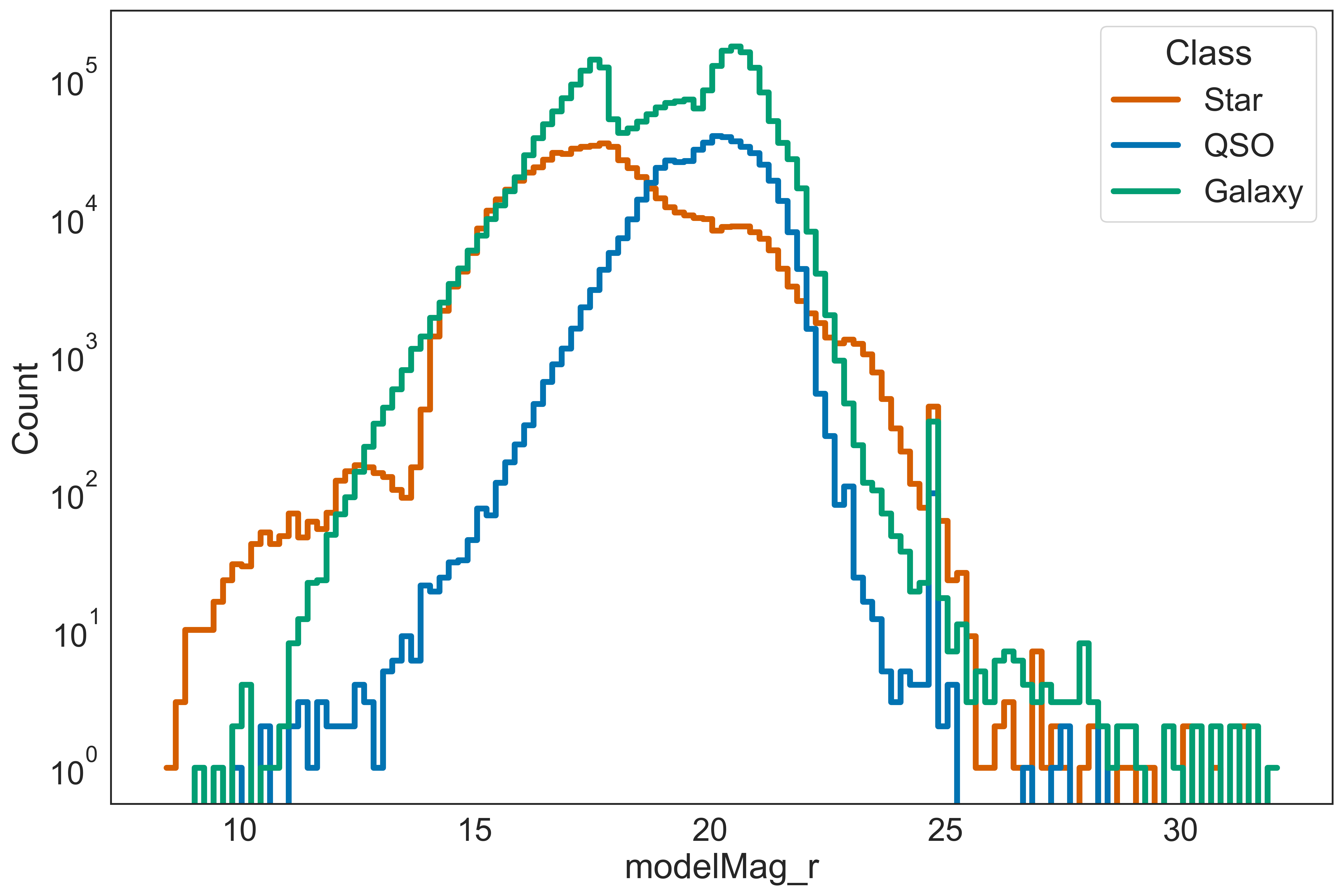}
  \caption{Histogram of the modelMag$\_$r for the extracted SDSS DR15 sources. Missing values are not included. Each source is colour-coded: orange for the star class; blue for the QSO class; and green the galaxy class.}
  \label{fig:modelmagr_dist}
\end{figure}

\begin{figure}[h!]
\centering
  \includegraphics[width=1\linewidth]{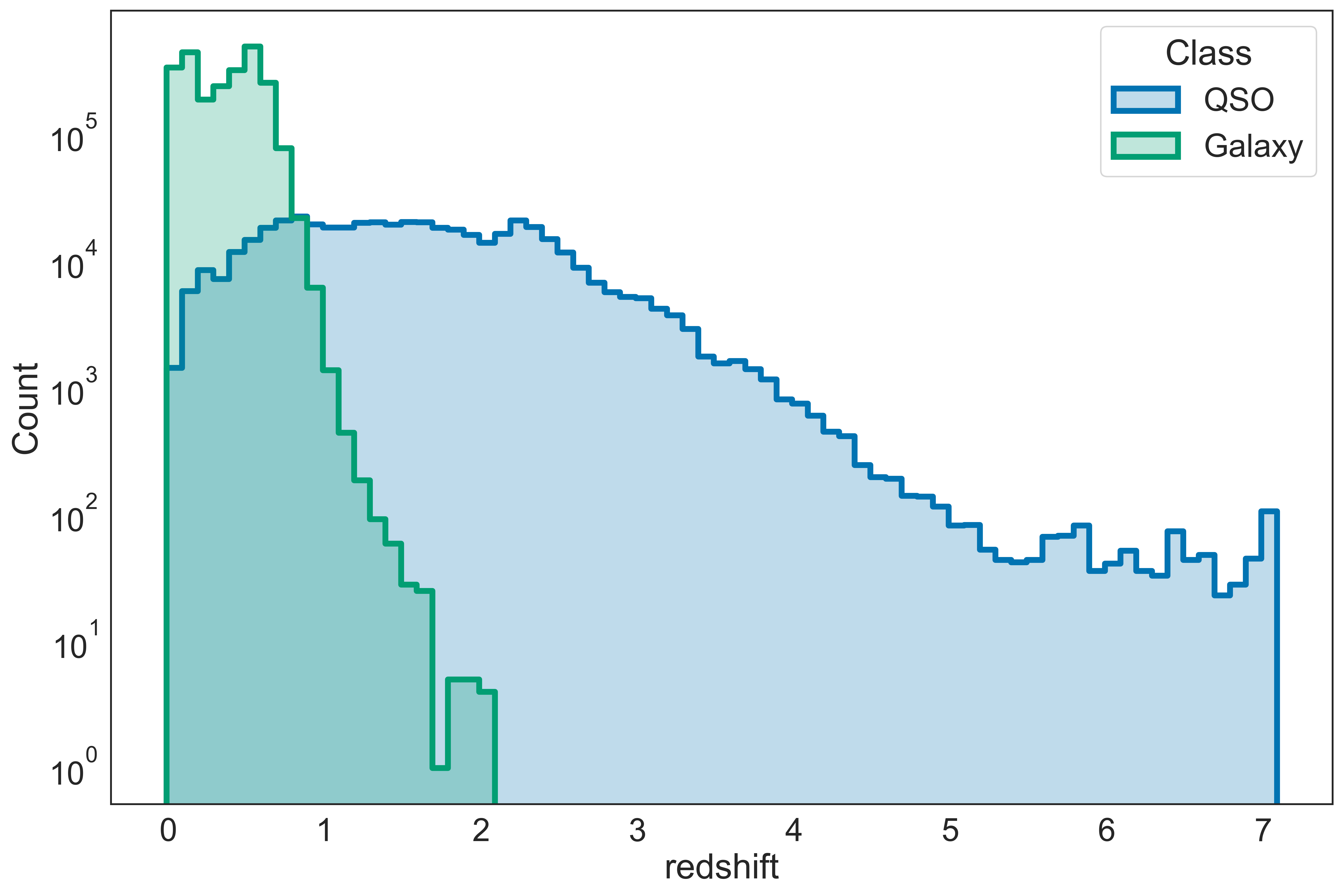}
  \caption{Histogram of the redshift distribution for the extracted SDSS DR15 sources. Each source is colour-coded: green for the galaxy class and blue  for the QSO class. The distribution for the star class is not represented since it is concentrated at z=0.}
  \label{fig:z_dist}
\end{figure}

 In this work an SQL query\footnote{\url{http://skyserver.sdss.org/CasJobs/}} was used to extract photometric data from SDSS Data Release 15 \citep{2019ApJS..240...23A} and WISE catalogues \citep{2010AJ....140.1868W}. The source ID was used to assure the cross-match between the catalogues. From the SDSS catalogue we extracted the psfMag, petroMag, cModelMag, and modelMag  for the five optical bands (u, g, r, i, z). The distribution of the modelMag magnitude for the r filter, by class, is represented in Fig. \ref{fig:modelmagr_dist}. The spectroscopic redshift distribution for the different class sources is represented in Fig. \ref{fig:z_dist}.
 
From the WISE catalogue, the four infrared bands (W1 3.4$\mu$, W2 4.6$\mu$, W3 12$\mu$, W4 22$\mu$) were extracted. In total, we extracted photometric data for 3,497,864 sources: 2,401,787 classified as galaxies, 473,954 as QSOs, and 599,177 as stars\footnote{SQL query can be found here: \url{https://github.com/pedro-acunha/SHEEP/blob/main/sql_query_casjobs}}. The SDSS spectroscopic classifications are template-based for all three classes; the galaxy class also contains some AGN whose detectable emission lines are consistent with Seyfert or LINER galaxies.

No pre-processing tasks were performed to remove missing values from the data set. All missing values are set as $-9999.0$, as in SDSS DR15. The properties of the missing values can be seen in Table \ref{table:data_missing}.
 
 \begin{table}[!h]
\caption{Fraction of the missing data in this work. Properties refers to the features being analysed, Count is the number of values missing for the correspondent variable, and Missing $\%$ is the percentage of the missing values compared with the total data. }             
\label{table:data_missing}      
\centering                          
\begin{tabular}{c c c}        
\hline\hline                
Properties & Count & Missing $\%$ \\    
\hline         
SDSS filter (u, g, r, i, z) & 171 & 0.005\\
WISE filter: W1 & 6 & 0.0001 \\
WISE filter: W2 & 106 & 0.003\\
WISE filter: W3 & 1394 & 0.04 \\
WISE filter: W4 & 331 & 0.01\\
\hline  \hline                                  
\end{tabular}
\end{table}

\section{Statistical metrics}
\label{metrics}
To evaluate the performance of any machine learning algorithm, either for classification or regression, statistical metrics are essential. In this section we give an overview of the different metrics used in this work. The statistical metrics for classification tasks vary between 0 and 1, where 1 is the perfect score. 

\subsection{Classification metrics}

For the classification tasks we used the following statistical metrics. Precision is the fraction of correct predictions for a given class compared to the overall number of predictions for that class,

\begin{equation}
\rm Precision = \frac{\rm TP}{\rm TP + \rm FP}\,,
\end{equation}

 \noindent where TP is the number of true positives in a given class and FP is the number of false positives in a given class. In astronomy this metric is also called "purity". 

Recall is the fraction of correct predictions for a given class, compared to the overall number of positives cases for that class,

\begin{equation}
\rm Recall = \frac{\rm TP}{\rm TP + \rm FN}\,,
\end{equation}
\noindent where FN is the number of false negatives in a given class. In other words, recall indicates the fraction of a class in the data set that the model correctly identifies. In astronomy this metric is also called "completeness".

The F1-score is the harmonic mean of the precision and recall,

\begin{equation}
{\rm F1{\text -}score} = 2\;{P \cdot R \over P+R}\,,
\end{equation}
\noindent where equal weight are given to precision and recall.

\vspace{2mm}

\subsection{Regression metrics}

For the regression tasks (i.e. redshift estimation) the following statistical metrics were adopted for consistency with \citealt{2020A&A...644A..31E}. The normalised median absolute deviation (NMAD) provides a measure of the variability in the sample,

\begin{equation}
    \mbox{NMAD} = 1.48 \cdot \mbox{median}\,\left(\frac{\,|z_{\rm pred}-z_{\rm ref}|\,}{1+z_{\rm ref}}\right)\,,
\end{equation}
\noindent where $z_{pred}$ is the predicted redshift and $z_{ref}$ is the ground-truth redshift given by the SDSS pipeline.

The fraction of catastrophic outliers, $f_{out}$, (e.g. \citealt{2010A&A...523A..31H}) is a quality control metric defined as

\begin{equation}
    \frac{|z_{\rm phot}-z_{\rm ref}|}{1+z_{\rm ref}} > 0.15.
\end{equation}

The bias in the predicted redshift can be estimated by studying the fluctuation of the predicted values,

\begin{equation}
    \mbox{bias} = \mbox{median}\left(\frac{z_{\rm phot}-z_{\rm ref}}{1+z_{\rm ref}}\right).
\end{equation}

\section{Proof of concept}
\label{POC}

Proof of concept (POC) tasks are essential to understand the viability of the classification and/or regression task. We designed three POC tasks: 
\begin{enumerate}
    \item Simulating the results obtained by \citealt{2020A&A...639A..84C}, using the same machine learning algorithm as the cited work (\texttt{Random Forest}\footnote{\url{https://scikit-learn.org/}} \citep{2001MachL..45....5B});
    
    \item Testing the gradient-boosting algorithm \texttt{XGBoost} \citep{2018arXiv180611248M} on the same data, for bench-marking;
    
    \item Exploring the impact of adding redshift as a feature to the SDSS and WISE photometry data by studying the impact in the classification statistical metrics.
\end{enumerate}

The same internal configuration of the classification algorithm (\texttt{Random Forest} and \texttt{XGBoost}) were used throughout this analysis. Fixing the hyper-parameters is important for an unbiased performance comparison. The classification metrics for the POC tasks are combined in Table \ref{table:clf_poc}. The column spec-z is binary, where the value No indicates that the spectroscopic redshift was not added to the data; the  value Yes indicates that it was added to the data.

\begin{table*}[!h]
\caption{Comparison of the classification metrics for each learning model in the POC tasks. For the reproduction and improvement of the results by \cite{2020A&A...639A..84C}, the spec-z, spectroscopic redshift, column is set as No; For the exploratory analysis of the inclusion of the spectroscopic redshift in the data, the spec-z column is set as Yes.}             
\label{table:clf_poc}      
\centering                          
\begin{tabular}{c c c c c c}        
\hline\hline                
spec-z & Algorithm & Accuracy & Precision & Recall & F1-score \\    
\hline                        
 No & Random-Forest (Clarke) & 0.983 & 0.981 & 0.967 & 0.974\\
 No & \texttt{XGBoost} & 0.987 & 0.984 & 0.977 & 0.980 \\
 Yes & Random-Forest & 0.989 & 0.991 & 0.979 & 0.984\\
 Yes &  \texttt{XGBoost} & 0.994  & 0.993 & 0.989 & 0.991\\
\hline  \hline                                  
\end{tabular}
\end{table*}

For the POC tasks (1) and (2), we used the same data and feature engineering as described in \cite{2020A&A...639A..84C}. We reproduced the results when using the \texttt{Random Forest} algorithm, achieving identical statistical metrics. With \texttt{XGBoost}, we obtained an increase of 0.004 in accuracy, 0.003 in precision, 0.01 in recall, and 0.006 in F1-score, compared to the metrics obtained using \texttt{Random Forest}.

In the third POC task we used the previously described photometric data, and the spectroscopic redshift was added. The same algorithms and internal configurations from the previous POC task were applied to the modified data. In the internal classification metrics analysis, \texttt{XGBoost} showed a slight improvement in the overall performance, compared with \texttt{Random Forest}, with an increase of 0.005 in accuracy, 0.002 in precision, 0.01 in recall, and 0.007 in F1-score. The overall increase in performance is consistent with that observed in the first POC task. To understand how the performance of the classification algorithms changes between the POC tasks, we compared the best algorithm from each task. An increase of 0.007 in accuracy, 0.009 in precision, 0.012 in recall, and 0.011 in F1-score was achieved with the addition of the redshift.

In summary, the main conclusions from the exploratory POC tasks are  that \texttt{XGBoost} outperforms \texttt{Random Forest} for the POC classification tasks  and  the inclusion of the spectroscopic redshift as a feature significantly increases the performance of both classification algorithms. To take advantage of the results from the third POC task, photo-z will be estimated and used in the classification tasks. The detailed description is found in Section \ref{photo_z}.

\section{The \texttt{SHEEP} pipeline}
\label{sheep}
We describe \texttt{SHEEP}\footnote{\url{https://github.com/pedro-acunha/SHEEP}}, a machine learning pipeline built to perform regression and classification tasks using tabular data. It is designed to perform  photo-z estimation and automatised source classification for tabular data. While in this work SDSS and WISE photometry are used, others types of tabular data are compatible with our architecture, such as radio fluxes. In the following subsubsections we present the general structure of our pipeline to allow reproducibility and further improvements. Since we are working with a large data set, only gradient boosting decision tree algorithms\footnote{An interesting benchmark study related to gradient boosting decision tree algorithms can be found in \cite{2018arXiv180904559A}} with GPU compatibility were explored. The Gradient platform\footnote{\url{https://gradient.run/}} was used to allow GPU acceleration with the NVIDIA RAPIDS framework \citep{2020arXiv200204803R}.

\subsection{Feature engineering}
\label{feat_eng}

 All unique permutations of broad-band colours were calculated from the following parameters: cModelMag; modelMag; psfMag; WISE; psfMag-WISE. Additional features using broad-band magnitudes were engineered as follows: psfMag-modelMag; psfMag-cmodelMag. The latter allows for the study of variance between different magnitudes measurements for the same filter, which can help the model to differentiate between compact and extended objects. Missing values for the magnitudes, colours, and engineered features were set as $-9999.0$. For the regression tasks the spectroscopic redshift was set as the target;for the classification tasks the source class was set as the target. Depending on the task being performed, additional targets may be defined, as explained in subsection \ref{learning_alg}.

Decision tree ensembles are not sensitive to the variance in the data set,
and thus the use of scaling does not affect the performance of the algorithm. For this reason we did not perform any feature scaling. 

\subsection{Learning algorithm}
\label{learning_alg}
The \texttt{SHEEP} pipeline is divided into two branches, a regression model that estimates the photo-z and a classification model to identify sources as galaxies, QSOs, and stars. We use the following algorithms as our base learners: \texttt{XGBoost}\footnote{\url{https://XGBoost.ai/}} \citep{2018arXiv180611248M}, \texttt{LightGBM}\footnote{\url{https://github.com/microsoft/LightGBM}} \citep{ke2017}, and \texttt{CatBoost}\footnote{\url{https://CatBoost.ai/}} \citep{2017arXiv170609516P}. While these gradient-boosting decision tree algorithms  have similarities, each one has unique advantages that can be leveraged for studies such as ours. \texttt{XGBoost} presents some innovative features, in particular the ability to deal with sparse data, which is particularly relevant for astronomy data sets. \texttt{LightGBM} introduces a histogram-based approach for continuous variables, which  improves training time, as well as a leaf-wise tree growth, while \texttt{XGBoost} uses a level-wise tree growth. \texttt{CatBoost} relies on an ordered boosting approach, preventing overfitting due to the use of residuals from previous fits, also called prediction shift. To make use of the different methodologies, we use ensemble methods to combine them: soft-voting for regression and hard-voting for classification. 

The base learners were trained and optimised using \texttt{FLAML} \citep{2019arXiv191104706W}, a Python library that explores the hyper-parameter space for each learner. \texttt{FLAML}\footnote{\url{https://microsoft.github.io/FLAML}} is built to provide cost-effective hyper-parameter optimisation. To ensure the generalisation of our model, we applied a data partitioning strategy called k-fold cross-validation (where k defines the number of partitions), where k was set as five. In resume, the full dataset is randomly split into k-folds (or parts) with similar size and with a balanced number of target classes. The model is trained using the k-1 folds and validated on the remaining fold, also known as out-of-fold (OOF). The model performance is computed averaging the predictions for k iterations. This procedure prevents overfitting by evaluating the model in k different validation sets, and  allows the computation of OOF predictions for the entire data set.

\begin{figure*}[h]
\centering
  \includegraphics[width=1\textwidth]{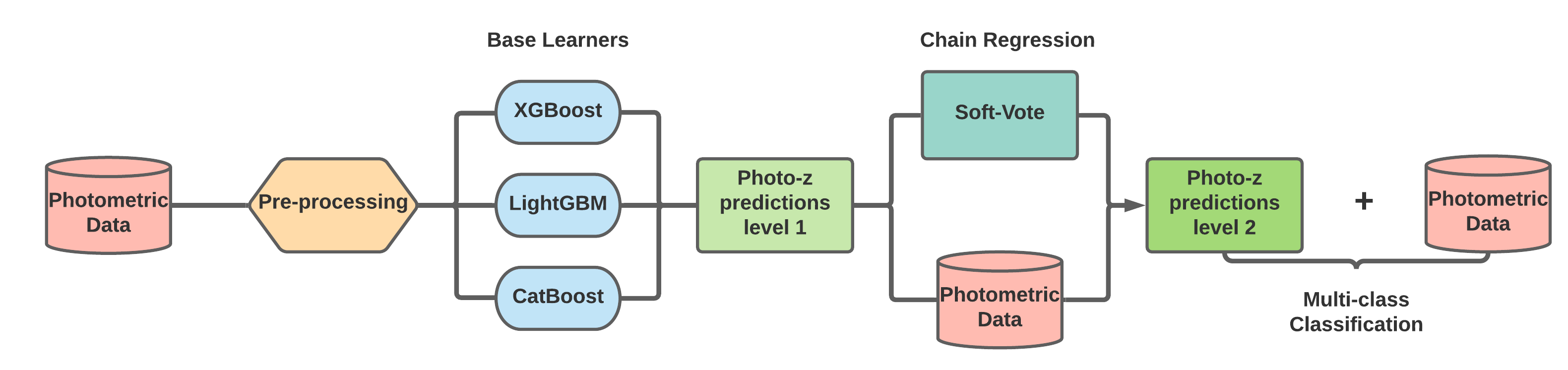}
  \caption{Flow chart describing the photo-z regression learning algorithm used in the \texttt{SHEEP} pipeline.}
  \label{fig:SHEEP_rgr}
\end{figure*}

\subsection{Photo-z estimation}
\label{photo_z}
Motivated by our finding that including redshift as a feature can improve classification accuracy, we implemented a photo-z prediction algorithm as part of our pipeline to predict the spectroscopic redshift of galaxies, QSOs, and stars from SDSS. In Fig. \ref{fig:SHEEP_rgr}, we show a flow chart describing the regression learning algorithm. After the feature engineering task, the data is split using a train-to-test ratio of 70:30. This split is motivated by the need to increase the training set, while still having a representative hold-out test set for model evaluation. The base learners are optimised and trained to calculate the first level for the photo-z predictions. 

Due to the limitations and similarities in the photometric data, the photo-z predictions will have catastrophic outliers. For improvement purposes, we added a second layer to the algorithm using a chain regressor approach, where the first-level predictions from the base learners are combined using a soft-vote methodology and included as a new feature in the data set. The final base learner models are trained on the new data set, providing the second-level predictions that are combined again by a soft-vote methodology to create the final photo-z predictions. These results are then inserted as a feature in the initial data set to be used in the classification task. The  first-level photo-z predictions, from the individual models and the combined predictions, are discarded.

\subsection{Classification}

\begin{figure*}[thb]
  \centering
  \includegraphics[width=0.7\linewidth]{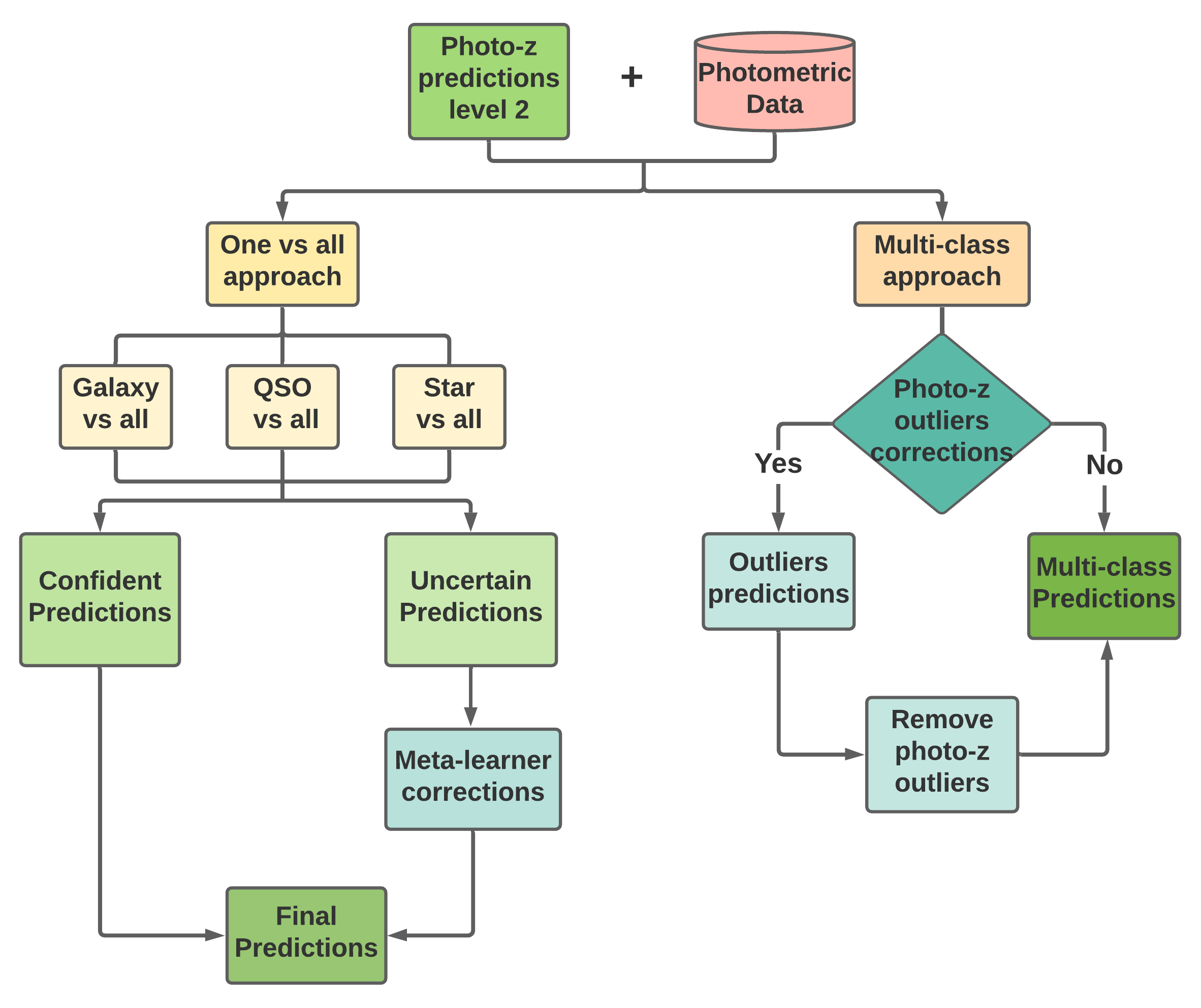}
  \caption{Flow chart describing the classification learning algorithm used in the \texttt{SHEEP} pipeline. }
  \label{fig:SHEEP_clf}
\end{figure*}

In Fig. \ref{fig:SHEEP_clf} we show a flow chart describing the classification learning algorithm. After the pre-processing task, the data is split using a train-to-test ratio of 50:50. The algorithm contains two approaches: multi-class and one versus all.

\subsubsection{Multi-class}
A model is trained to output multiple class predictions, while allowing the removal of potential catastrophic outliers from the photo-z estimation. We refer to this option as the  \texttt{monitoring model}, and it must be activated in the early stages of the pipeline. This model is a binary classification model trained and optimised, using \texttt{FLAML} library, to identify potential catastrophic outliers \citep[e.g.][]{2022ApJ...928....6S}. If the \texttt{monitoring model} is requested, the potential catastrophic outlier photo-z estimations are removed. The training data remains unchanged independently of the activation of the \texttt{monitoring model}. Finally, a classification model is used to predict the testing data. The final classification model is the same for both approaches.

\subsubsection{One versus all}
The one versus all approach allows the classification problem to be divided into multiple binary classification problems: galaxy versus all; QSO versus all; star versus all. The three models produced will be specialised for each source, allowing a grasp of deeper connections in the data. The one versus all approach presented here was inspired by \cite{2020A&A...633A.154L}. We added a more detailed exploration of this concept with alternative solutions and analysis. The most interesting, and also overlooked, aspect of this approach is that sources can be classified with only one class, for example  the source is classified as a galaxy in the galaxy versus all model, and not as a QSO or star in the others.

When only one class label is assigned to a source from the three binary classification models, we denominate it as a \textit{confident prediction}. If the source is assigned multiple labels or none, we denominate it as an \textit{uncertain prediction}. The uncertain predictions can highlight interesting study cases since they can originate from (i) a lack of generalisation from the model itself  or  (ii) a similarity in the photometry data confusing the models. Case (ii) can be physically motivated (e.g. rare objects, transitional sources).

Two solutions were designed to deal with the uncertain predictions: multi-class and meta-learner.
\begin{itemize}
    \vspace{2mm}
    \item The multi-class model is built with the assumption that most of the incorrect predictions are due to limitations in the pipeline models. The training data is used by our base learner algorithms to perform a multi-class classification using the uncertain predictions as the test data. The main idea is make use of the generalisation strength to provide a final predicted label for the test data, while allowing parameter tuning for the classification probability threshold. The decision tree ensemble used for the uncertain prediction corrections are the same as previously computed in the main branch multi-class approach. To avoid the introduction of biases from separate approaches and to fairly compare the efficiency provided by two different and independent methodologies, this solution was not implemented.
    \vspace{2mm}
    \item In the meta-learner approach, generalised stacking \citep{wolpert} is an ensemble method that takes multiple weak learners and combines them, either by building a new predictive meta-model or by combining the weak learner predictions. As in the multi-class solution, we assume that the uncertain predictions are caused by a  lack of generalisation. The generalised stacking with meta-learner method is used to label the uncertain predictions by stacking the predictions from the base learners, for each of the classes, in the one versus all starting step. A meta-learner is trained with the correct prediction classification probabilities, and then used to make the final predictions for the uncertain predictions. This process allows the meta-learner to learn the inner structure of the predictions and learn how to improve them, correcting potential misclassifications made by the initial stacking. By setting a probability threshold, the number of misclassifications can be controlled and further studied. In this work the task-oriented \texttt{AutoML} from the \texttt{FLAML} library was used to select the most accurate algorithms from our base learners to be used as the meta-learner. The \texttt{LightGBM} algorithm  was selected and trained using the predictions probabilities from the confident predictions given by the base learners.
\end{itemize}

\subsection{Target variable}

Inside the \texttt{SHEEP} pipeline the target feature is set dynamically, depending on the task. For photo-z regression the target is continuous and is set as $1 + z$. For the classification tasks the target feature is always discrete. For the multi-class models the target variable is set to 0 for galaxies, 1 for QSO, and 2 for stars. In the one versus all approach the class to be identified (galaxy, QSO, or star) is assigned a target value of 0, while  the remaining sources are assigned a target value of  1. For the outlier detection the target feature is set as 0 for `Not outlier' and 1 for `Outlier'.

\subsection{Feature selection and importance}

\begin{figure*}[!h]
\centering
\subfloat[Multi-class]{\label{a} \includegraphics[width=.47\linewidth]{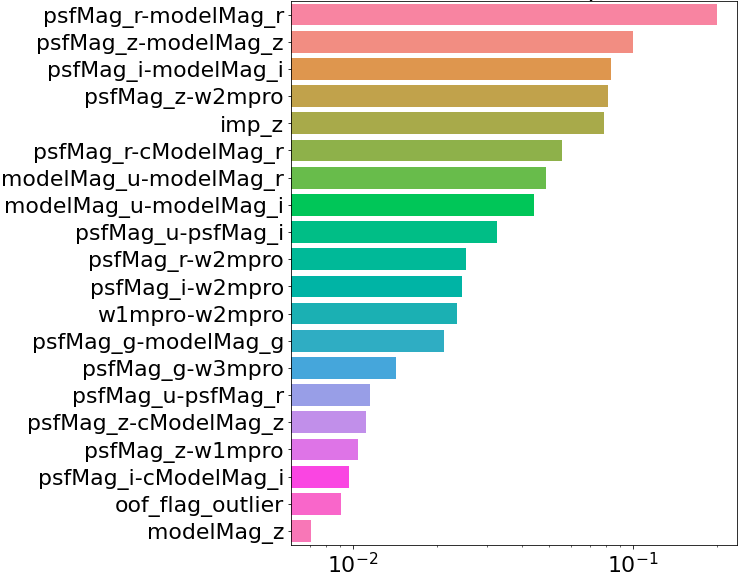}}\hfill
\subfloat[Galaxy vs all]{\label{b}\includegraphics[width=.46\linewidth]{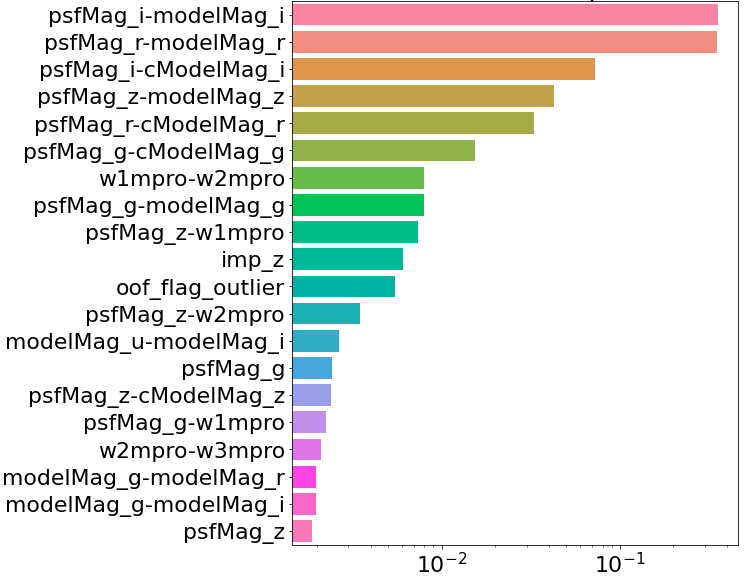}}\par
\subfloat[QSO vs all]{\label{c}\includegraphics[width=.5\linewidth]{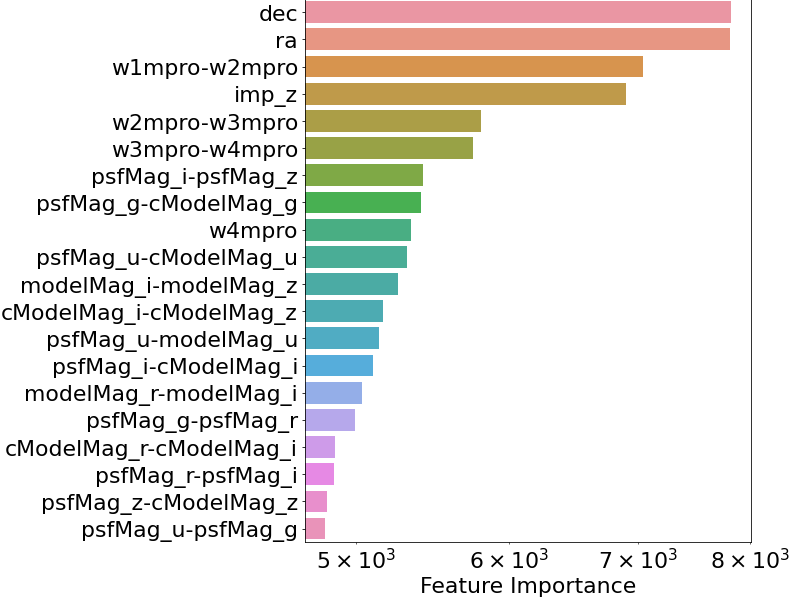}}\hfill
\subfloat[Star vs all]{\label{d}\includegraphics[width=.5\linewidth]{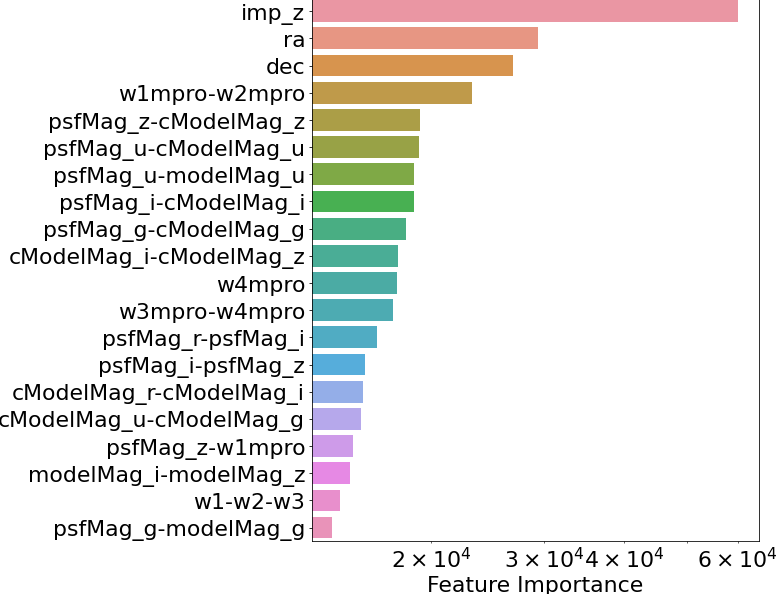}}
\caption{Importance values of top 20 features  for the best models in the multi-class and one vs all approach: (a) and (b) \texttt{XGBoost}; (c) and (d) \texttt{LightGBM}. The  importance values (x-axis) are presented in logarithmic scale for ease of visualisation. The x-axis scale also varies based on the algorithm being used. The feature names are shown on the y-axis. The variable \texttt{imp$\_$z} is the final photo-z prediction value and \texttt{oof$\_$flag$\_$outlier} is the binary variable that identifies the out-of-fold photo-z catastrophic outliers.}
\label{fig:feat_import}
\end{figure*}

For model explainability, decision tree-based algorithms provide the user with feature importance values. This is relevant for understanding the  importance of the individual features used in the model. However, different learning algorithms have a unique way of calculating the feature importance. For example, in the \texttt{XGBoostClassifier} the default option is the gain, which represents the improvement in accuracy that each feature is able to provide. In Fig. \ref{fig:feat_import} the top 20 features are shown for the multi-class learning algorithm. This analysis reveals some intriguing results, such as the presence of the \texttt{oof$\_$flag$\_$outlier} binary feature that denotes whether the photo-z is expected to be an outlier. This is also seen for \texttt{LightGBM} and \texttt{CatBoost}, even though the feature importance value varies between the learning algorithms. 

The presence of the photometric redshift in Fig. \ref{fig:feat_import} is physically explainable (see Section \ref{POC}). For example, individual stars are only detected at redshifts very close to zero, and QSOs have a bias towards higher redshifts than galaxies. The learning algorithm is able to pick up those relationships and correlate them to physical features, such as psfMag$\_$r-modelMag$\_$r and w1mpro-w2mpro.

A/B tests were conducted to understand whether the coordinates of each source, right ascension (RA) and declination (Dec), would help the learning algorithm. No significant improvement was observed for the multi-class approach. These features were used in the one versus all approach.

In Fig. \ref{fig:feat_import} the top 20 features are represented for the three one versus all models. As expected, the photo-z related features, \texttt{imp$\_$z} and \texttt{oof$\_$flag$\_$outlier}, and the photometric colours are among the most important features, independently of the task. Infrared information, either magnitudes or colours, are also relevant, in particular for QSO and star classification. 

Interestingly, the RA and Dec values are among the most important features in the QSO versus all and star versus all \texttt{LightGBM} models. Stars are distributed closer to the Galaxy disk, while QSOs are easier to find away from the dusty Milky Way disk. Since the relative density varies across the SDSS survey area, the models can take into account the local density of these classes when estimating the probability that an unresolved source is a star or a QSO. 

\section{Results and discussion}
\label{results}
\subsection{Photometric redshifts}
\label{results_photo_z}
 The photo-z regression learning algorithm provides  predictions at two levels. For the first level the photo-z is predicted using photometric data, as described in Section \ref{feat_eng}. The metrics for the first- and second-level predictions are shown in Table \ref{table:rgr_specz}. We also extracted SDSS photometric redshift estimations, derived by robust fit to the nearest neighbours in a reference set, for 2,473,520 galaxy sources and added them for comparison.

\begin{table*}[htp]
\caption{\textit{Top rows:} Regression metrics for photo-z predictions level one and two. The metrics for \texttt{XGBoost}, \texttt{CatBoost} and \texttt{LightGBM} are representative of the testing data.  Ensemble is the average of the predictions given by the base learners, with models built in the k-fold procedure using the training data. 
\textit{Bottom rows:} Regression metrics for the SDSS photo-z prediction and this work, only for galaxy sources.}
\label{table:rgr_specz}      
\centering          
\begin{tabular}{c c c c c c c c c c c }      
\hline\hline    
\textbf{Algorithm} & \multicolumn{4}{c}{\textbf{Level one}} & \multicolumn{4}{c}{\textbf{Level two}} \\
\cline{2-5} \cline{5-9}
 & Bias & NMAD & $R^2$ & Outlier fraction & Bias & NMAD & $R^2$ &  Outlier fraction \\ 
  {\texttt{XGBoost}} &  0.015 & 0.021 & 0.899 & 0.0376 & 0.013 & 0.018 & 0.909 & 0.0290\\
 {\texttt{CatBoost}} & 0.015 & 0.021 & 0.905 & 0.0347 & 0.013 & 0.018 & 0.914 & 0.0272 \\ 
  {\texttt{LightGBM}}&  0.015 & 0.021 & 0.903 & 0.0344 & 0.013 & 0.018 & 0.915 & 0.0270 \\ 
  Ensemble & 0.014 & 0.019 & 0.906 & 0.0342 & 0.0124 & 0.018 & 0.916 & 0.0265 \\
  \hline
  SDSS (galaxies only) & - & - & - & - & 0.014 & 0.020 & 0.714 & 0.0226\\
  This work (galaxies only) & - & - & - & - & 0.013 & 0.019 & 0.912 & 0.0081\\
\hline \hline            
\end{tabular}
\end{table*}

The regression metrics for level two are significantly improved, when compared with the predictions for level one. The addition of the chain regressor approach helps the new model to recognise and correct some of the inconsistencies in the level one predictions. In a comparison of  our work with the SDSS photo-z predictions, we outperform the SDSS algorithm.

\begin{figure*}
    \centering
    \includegraphics[width=\linewidth]{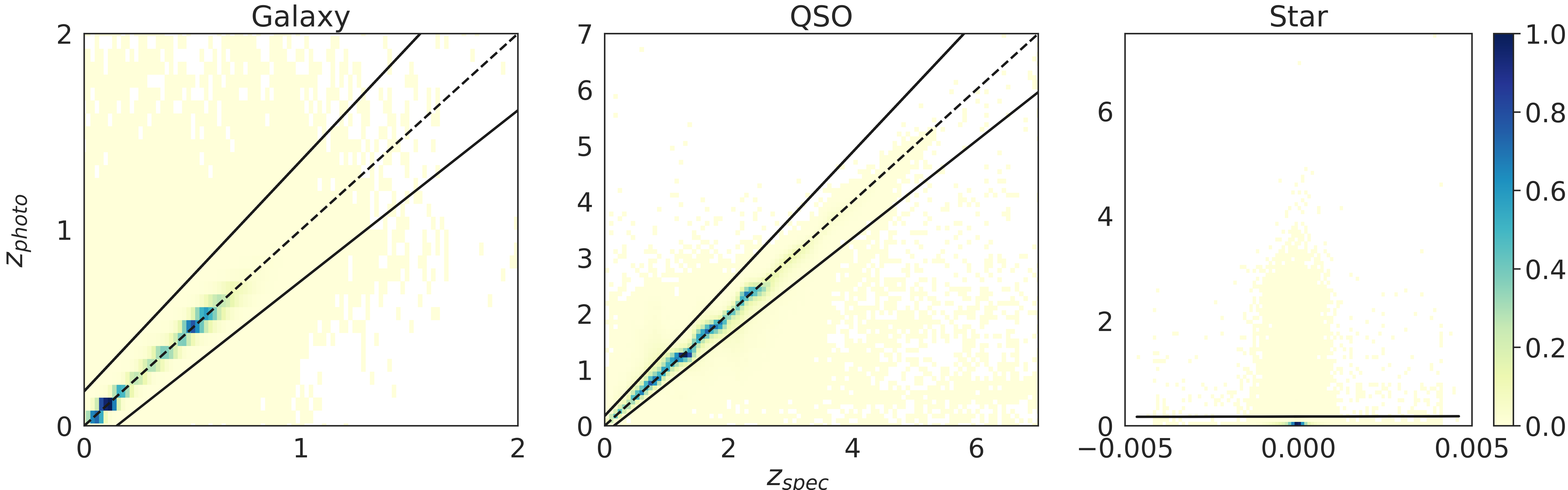}
    \hfill
    \caption{Distribution of the spectroscopic redshifts ($z_{spec}$) as a function of the predicted photometric redshifts ($z_{photo}$). The centre dashed line indicates the $y=x$, while the other solid lines delimit the outlier region. In the star class plot, only the outlier line is represented as a solid line. The colour bar is normalised, where blue represent the denser regions and yellow represents the rarer regions.}
    \label{fig:plot_2dhist}
\end{figure*}

\begin{figure}
    \centering
    \includegraphics[width=\linewidth]{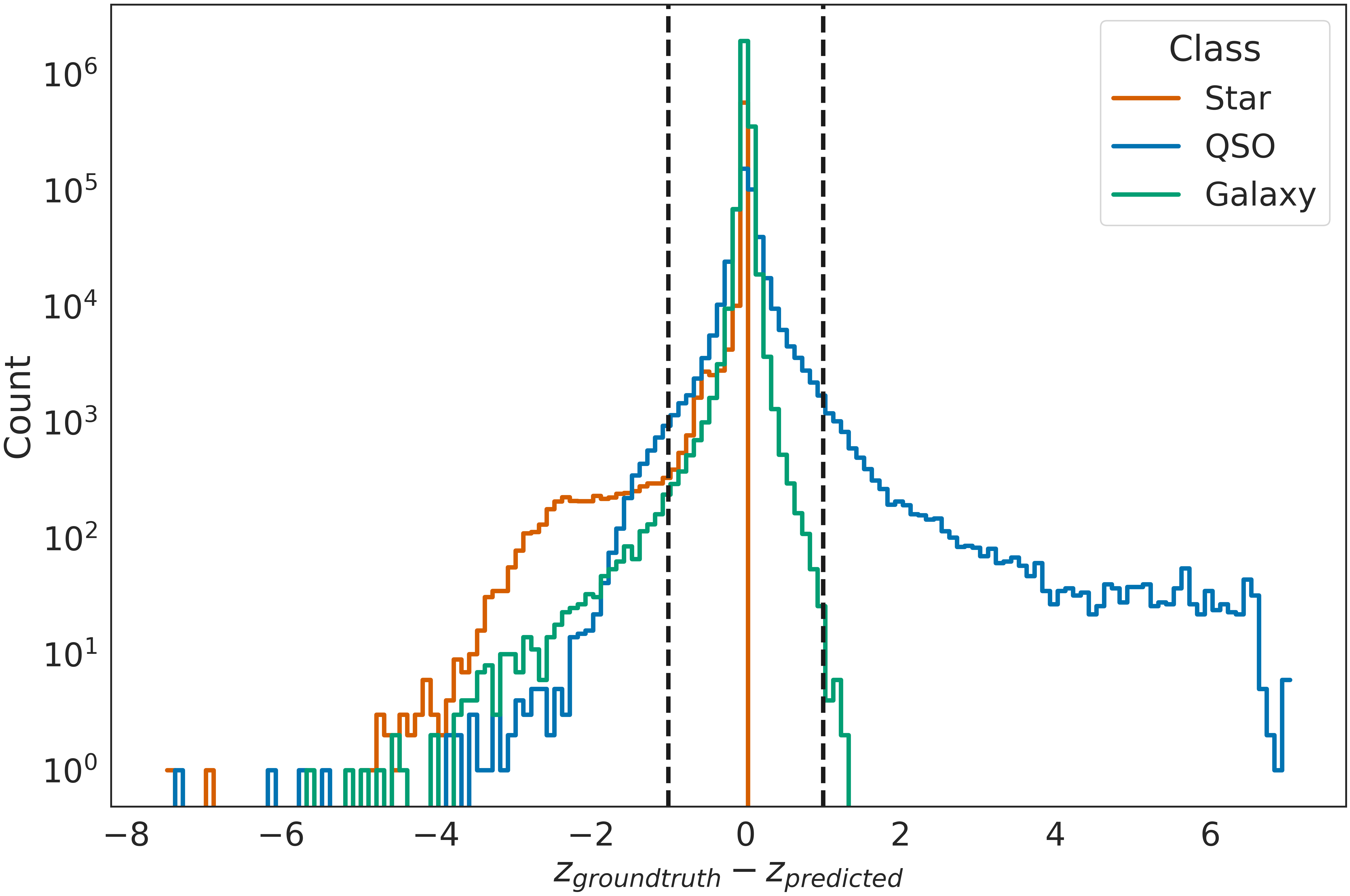}
    \hfill
    \caption{Distribution of the difference between the spectroscopic redshift ($z_{ground truth}$) provided by SDSS spectroscopy and the photometric redshift predicted in our work ($z_{predicted}$). The histogram is colour-coded by spectroscopic label: orange for stars, blue for QSOs, and green for galaxies. The black dashed lines delimit the area where the difference between redshifts is $1$ or $-1$.}
    \label{fig:photo_z}
\end{figure}

In Fig. \ref{fig:plot_2dhist}, the predicted photo-z for each class is plotted in comparison to the ground-truth spectroscopic redshift. For the galaxy and QSO classes, a linear relation is clearly verified with a greater number of sources lying close to the 1:1 relation. For the star class, a more dispersed distribution is observed due to difficulties in distinguishing stars from QSOs. 
The distribution of the difference between the spectroscopic redshift and the photo-z is represented in Fig. \ref{fig:photo_z}. The galaxy and star sources present a peak around value zero; the catastrophic outliers are due to overestimations from the prediction model. For QSO sources, over- and underestimations are detected, with higher distribution towards underestimation. 

\subsection{Galaxy, QSO, star classifications}

We present the results from our unique architecture (see Fig. \ref{fig:SHEEP_clf}). The photo-z predictions were added to the photometric data and included in the \texttt{SHEEP} pipeline. 

\subsubsection{Multi-class approach}
\label{multiclass_approach}

In Table \ref{table:clf_all} the final classification metrics are represented for the case where the photo-z catastrophic outliers corrections is not chosen. The best model performing algorithm was \texttt{XGBoost}. The impact of quality for the photo-$z$ predictions, using the \texttt{XGBoost} algorithm, are represented in Table \ref{table:clf_cat_out}. The classification metrics are  improved when potential catastrophic outliers are removed from the data set. Nonetheless, the improvement is also correlated to the efficiency of the \texttt{monitoring model}. In our case we obtained the following  F1-score, 0.56 for catastrophic outliers, and 0.99 for non outliers. Only 57$\%$ of the catastrophic outliers were detected and removed from our sample. By improving the detection and extraction these outliers the model can be easily improved either by removing them or correcting them.

\begin{table*}[!h]
\caption{Classification metrics for the \texttt{XGBoost} classification model in the multi-class approach, with the \texttt{monitoring model} for the photo-$z$ predictions. The impact of the removal of potential catastrophic outliers is studied.}             
\label{table:clf_cat_out}      
\centering                          
\begin{tabular}{c c c c c c}       
\hline\hline                 
Drop outliers & Accuracy & Precision & Recall & F1-score \\ 
\hline                        
 No & 0.988 & 0.985 & 0.979 & 0.982\\
 Yes & 0.989 & 0.988 & 0.981 & 0.984\\ \hline 
\hline                                  
\end{tabular}
\end{table*}

\subsubsection{One versus all approach}
\label{one_vs_all}

The classification metrics for the individual learning algorithms in the one versus all approach are compiled in Table \ref{table:clf_all}. All base learners perform similarly well. It is important to remember that a small increase of 0.002 in recall is still relevant when dealing with big data. For example, in the QSO versus all, this increase in recall results in more 1,800 sources being correctly classified. This is important to improve our machine learning models and reduce further human analysis. The \texttt{XGBoost} algorithm performed better for galaxy data, while the \texttt{LightGBM} algorithm performed better with the QSO and star data. 

\begin{table*}[!h]
\caption{Classification metrics for the \texttt{XGBoost}, \texttt{LightGBM}, and \texttt{CatBoost} models for the multi-class and one versus all approaches. The metric values shown are calculated over all classes.}   
\label{table:clf_all}      
\centering                          
\begin{tabular}{c c c c c c c }        
\hline\hline                 
Approach & Class & Algorithm & Accuracy & Precision & Recall & F1-score \\   
\hline                       
&  &\texttt{XGBoost}  & 0.988 & 0.985 & 0.978 & 0.981  \\
Multi-class & Multi & \texttt{CatBoost} & 0.986 & 0.983 & 0.976 & 0.979 \\
&  & \texttt{LightGBM} & 0.987 & 0.984 & 0.978 & 0.981 \\
\cline{1-7} 
 & & \texttt{XGBoost}  & 0.989 & 0.989 & 0.986 & 0.988 \\
& Galaxy & \texttt{LightGBM} & 0.989 & 0.989 & 0.986 & 0.987  \\
& & \texttt{CatBoost} & 0.988 & 0.988 & 0.984 & 0.986 \\
 
\cline{2-7} 
& & \texttt{XGBoost}  & 0.991 & 0.985 & 0.978 & 0.981  \\
One vs all & QSO & \texttt{LightGBM} & 0.991 & 0.985 & 0.978 & 0.982 \\
& & \texttt{CatBoost} & 0.990 & 0.983 & 0.976 & 0.980  \\
\cline{2-7} 
& & \texttt{XGBoost}  & 0.995 & 0.993 & 0.988 & 0.991\\
& Star & \texttt{LightGBM} & 0.995 & 0.993 & 0.989 & 0.991  \\
& & \texttt{CatBoost} & 0.994 & 0.992 & 0.987 & 0.990\\
\hline    \hline  
\end{tabular}
\end{table*}

To compare directly the predictions made with the main approaches, only using the sources with one class designation, we computed the F1-scores:  multi-class: 0.98389;  one versus all: 0.98398. The two methodologies have similar classification metrics, but the one versus all approach is slightly better.

There were 5,285 uncertain predictions from the total test data set of 1,748,932 sources, representing around 0.3$\%$ of the test data. A total of  801 sources were classified as galaxy and QSO; 814 sources as galaxy and star; 288 sources as QSO and star; and 3,382 sources had no class. The meta-learner algorithm, \texttt{LightGBM}, predicted the class sources for the uncertain predictions with the following metrics:  Precision: 0.585;  Recall: 0.592;  F1-score: 0.583.

\subsection{Comparison with similar studies}
\begin{figure*}[h]
  \centering
  \includegraphics[width=\linewidth]{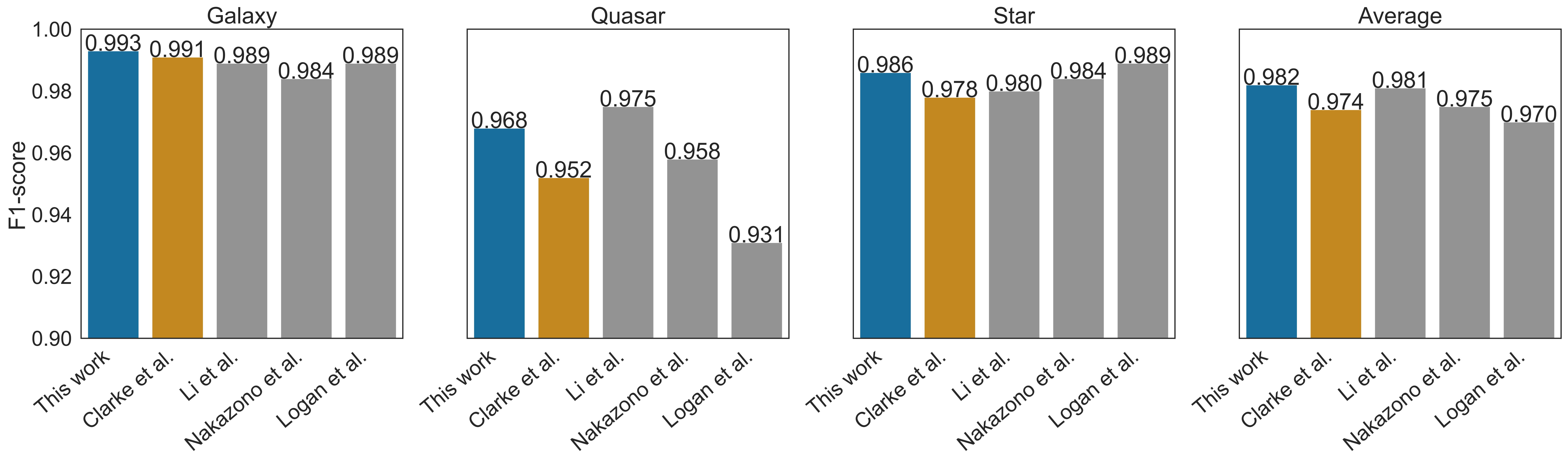}
  \caption{F1-score metrics from this work (one vs all with meta-learner corrections), \cite{2020A&A...639A..84C}, \cite{2021MNRAS.506.1651L}, \cite{2021MNRAS.507.5847N}, and \cite{2020A&A...633A.154L}. The first three plots represent the metrics for each individual source class: galaxy, QSO, and star. The fourth and fifth plots represent the average F1-score for each work.}
  \label{fig:compare}
\end{figure*}

This study was designed to allow a direct comparison with the random forest-based classification methodology of \cite{2020A&A...639A..84C}. A histogram combining the F1-scores for the different classes are shown in Fig. \ref{fig:compare}. The precision, recall, and F1-score metrics are given in Table \ref{table:final}. Our method outperforms the approach of \cite{2020A&A...639A..84C}, in terms of the F1-score, for all three classes. In general, we obtain improved precision and recall metrics values compared to those obtained by \cite{2020A&A...639A..84C}. Two exceptions are the precision for the star class, which is marginally below that of \cite{2020A&A...639A..84C}, and the recall for the galaxy class, where a similar result was obtained.

\begin{table*}[htp]
\caption{Comparison of the classification metrics from our methodology, multi-class and one versus all, with \cite{2020A&A...639A..84C} .}
\label{table:final}      
\centering          
\begin{tabular}{c c l l l c l l l c l l l }     
\hline\hline    
\textbf{Work} & \multicolumn{3}{c}{\textbf{Precision}} & \multicolumn{3}{c}{\textbf{Recall}} & \multicolumn{3}{c}{\textbf{F1-score}}\\ \cline{2-4} \cline{5-7} \cline{8-10}
 & Galaxies & Quasars & Stars & Galaxies & Quasars & Stars & Galaxies & Quasars & Stars \\ 
  {Multi-class approach} & 0.989 & 0.975 & 0.990 &  0.995 & 0.960 & 0.980 & 0.992 & 0.967 & 0.985\\
 {One vs all approach} & 0.990 & 0.975 & 0.990 & 0.995 & 0.961 & 0.981 & 0.993 & 0.968 & 0.986\\ 
  {Clarke et al. (2020)}& 0.987 & 0.961& 0.991& 0.995 &0.944 &0.965& 0.991 &0.952 &0.978 \\ 
\hline \hline            
\end{tabular}
\end{table*}

Our classification method also performs well when compared to other similar studies \citep[][see Fig. \ref{fig:compare}]{2020A&A...633A.154L, 2021MNRAS.506.1651L, 2021MNRAS.507.5847N}, although a direct comparison is not possible since significantly different data sets were used.

\section{Conclusions and final remarks}
\label{conclusions}
We have described \texttt{SHEEP}, a machine learning pipeline for the classification of galaxies, QSOs, and stars using tabular photometric data, while accepting sparse data. It starts by estimating the photo-z and adding it as a feature for the training tasks. The pipeline enables two types of approaches, multi-class with photo-z outlier detection and one versus all with meta-learner correction, to combine the outputs from gradient boosting decision trees into a single stronger classifier.

From our best approach, one versus all with meta-learner corrections for the uncertain predictions, we obtained a precision of 0.985, a recall of 0.979, and  an F1-score of 0.982, using photometric data from SDSS and WISE and the predicted photo-z. Our F1-scores are significantly higher than those obtain with the \texttt{Random Forest} method by \cite{2020A&A...639A..84C}. For the photo-z estimation, we obtained a NMAD of 0.010, a bias of 0.007, a catastrophic outlier fraction of 0.02, and an $R^2$ score of 0.916.

In future work we will adapt the SHEEP pipeline for application to data sets from upcoming surveys such as {\it Euclid}, where machine learning techniques will be indispensable due to the expected very large volumes of data. The inclusion of additional data will likely lead to further improvement in the classification performance of SHEEP, including (i) wider wavelength coverage (e.g. near-infrared photometry from {\it Euclid}); (ii) narrower band filters (e.g. from J-PAS); 
(iii) the introduction of morphological information from images; and (iv) the inclusion of spectral information. In addition, an exploration of the application of active learning with a hybrid methodology is deferred to a future publication (Cunha et al., in prep.).

\begin{acknowledgements}
PACC dedicates this work in memory of Prof. Eduardo Pereira. The authors thank Ana Afonso and Tom Scott for their comments and suggestions. PACC acknowledges financial support by Centro de Astrofísica da Universidade do Porto through grant CIAAUP-12/2021-BI-D from UIDB/04434/2020. AH acknowledges support from NVIDIA through an NVIDIA Academic Hardware Grant Award. AH acknowledges  financial support by Fundação para a Ciência e
a Tecnologia (FCT) through grants UID/FIS/04434/2019,
UIDB/04434/2020, UIDP/04434/2020 and PTDC/FISAST
/29245/2017, and an FCT-CAPES Transnational Cooperation
Project. Funding for the Sloan Digital Sky Survey IV has been provided
by the Alfred P. Sloan Foundation, the U.S. Department of Energy Office of
Science, and the Participating Institutions. SDSS-IV acknowledges support and
resources from the Center for High-Performance Computing at the University
of Utah. The SDSS website is http://www.sdss.org/. SDSS-IV is managed
by the Astrophysical Research Consortium for the Participating Institutions
of the SDSS Collaboration including the Brazilian Participation Group, the
Carnegie Institution for Science, Carnegie Mellon University, the Chilean
Participation Group, the French Participation Group, Harvard-Smithsonian
Center for Astrophysics, Instituto de Astrofísica de Canarias, The Johns
Hopkins University, Kavli Institute for the Physics and Mathematics of the
Universe (IPMU)/University of Tokyo, Lawrence Berkeley National Laboratory,
Leibniz Institut für Astrophysik Potsdam (AIP), Max-Planck-Institut für
Astronomie (MPIA Heidelberg), Max- Planck-Institut für Astrophysik (MPA
Garching), Max-Planck-Institut für Extraterrestrische Physik (MPE), National
Astronomical Observatories of China, New Mexico State University, New
York University, University of Notre Dame, Observatário Nacional/MCTI, The
Ohio State University, Pennsylvania State University, Shanghai Astronomical
Observatory, United Kingdom Participation Group, Universidad Nacional
Autónoma de México, University of Arizona, University of Colorado Boulder,
University of Oxford, University of Portsmouth, University of Utah, University
of Virginia, University of Washington, University of Wisconsin, Vanderbilt
University, and Yale University.
\end{acknowledgements}

%
%
\bibliographystyle{aa} 

\end{document}